\documentclass[preprint,showpacs,preprintnumbers,amsmath,amssymb]{revtex4}
\usepackage{graphicx}% Include figure files
\usepackage{dcolumn}% Align table columns on decimal point
\begin{document}

%\preprint{APS/123-QED}

\title{New mechanism of solution of the $kT$-problem in magnetobiology}% Force line breaks with \\

\author{Zakirjon Kanokov$^{1,2*}$, J\"{u}rn W. P. Schmelzer$^{1,3}$,
 Avazbek K. Nasirov$^{1,4}$}
\affiliation{$^1$Bogoliubov Laboratory of Theoretical Physics, Joint Institute
for Nuclear Research, Dubna, Russia}\email{zokirjon@yandex.ru}
\affiliation{$^2$Faculty of Physics, M. Ulugbek National University of Uzbekistan,
Tashkent, Uzbekistan}
\affiliation{$^3$Institut f\"ur Physik, Universit\"at Rostock,  Rostock, Germany}
\affiliation{$^4$Institute of Nuclear Physics,Tashkent,Uzbekistan}

\date{\today}% It is always \today, today,
\begin{abstract}
The effect of ultralow-frequency or static magnetic and electric fields on
biological processes is of huge interest  for researchers due to the resonant
change of the intensity of biochemical reactions although the energy in such
fields is small in comparison with the characteristic energy $k_BT$ of the
chemical reactions. In the present work a simplified model to study the effect
of the weak magnetic and electrical fields on fluctuation of the random ionic
currents in blood and to solve the $k_BT$ problem in magnetobiology is suggested.
The analytic expression for the kinetic energy of the molecules
dissolved in certain liquid media is obtained. The values of the magnetic field
leading to resonant effects in capillaries are estimated.  The numerical
estimates showed that the resonant values of the energy of molecular in the
capillaries and aorta are different. These estimates prove that under identical
conditions a molecule of the aorta gets $10^{-9}$ times less energy than the
molecules in blood capillaries. So the capillaries are very sensitive to the
resonant effect, with an approach to the resonant value of the magnetic field
strength, the average energy of the molecule localized in the capillary is
increased by several orders of magnitude as compared to its thermal energy,
this value of the energy is sufficient for the deterioration of the chemical bonds.
 \end{abstract}

\maketitle

\section{\label{Introduc}Introduction}

One of key problems magnetobiology is the explanation of the mechanism of
influence of weak magnetic fields on biological objects. The experimental
data show that influence of weak magnetic fields on the test-system
(both of animal \cite{Koch,Harland,Moses} %[1-3]
and botanical \cite{Bersani,Lednev} %[4,5]
origin) are realized by the same mechanism. As a consequence, similar effects
have to be expected to be of significance for human beings as well. A large
spectrum of data is known \cite{Ptitsyna,Binhi}, %[6,7]
accumulated in biology, biophysics, ecology, and medicine showing that all
ranges of the spectrum of electromagnetic radiation may influence health
and working ability of people.

The human organs are not capable to feel physically an
electromagnetic field surrounding a human's body; however, it may
cause the decrease of immunity and working ability of people, under
its influence syndromes of  chronic weariness may develop, and the
risk of diseases increases. The action of electromagnetic radiation
on children, teenagers, pregnant  women and persons with weakened
health is especially dangerous \cite{Ptitsyna}.%[6].
The negative influence of electromagnetic fields on human beings and other
biological objects is directly proportional to the field intensity and
irradiation time. Hereby a negative effect of the electromagnetic field
appears already at field strengths equal to 1000 V/m.  In particular, the
endocrine system, metabolic processes, functioning of head and a spinal
cord etc. of people may be affected \cite{Ptitsyna}.%[6]

The absence of a theoretical explanation of the mechanism of action
of weak magnetic fields on biological objects is connected mainly
with the so-called $k_BT$ problem. Here $k_B$ is the Boltzmann constant
and $T$ is the temperature of the medium. As it is noted in the review
\cite{Binhi}, %[7],
 at present does not exist a comprehensive understanding from the
 point of view of physics how weak low-frequency magnetic fields may affect
 living systems. In particular, it is not clear how low-frequency
 weak magnetic fields may lead to the resonant change of the rate of
 biochemical reactions although the impact energy is by ten orders of
 magnitude less then $k_BT$.

At the same time, up to now there is no theory in the framework of
the general physical concepts underlying magnetobiology and
heliobiology, i.e. in the field of a science where effects of the
weak and super-weak magnetic fields are studied, giving an answer to
these questions. There are even no qualitative theoretical models
explaining the interaction mechanisms of fields with biological
objects. From the  point of view of physics, this situation is
connected with complexity of the macroscopic open systems when the
concepts of physics, biology, and chemistry are applied. This
complexity is caused by the fact that being macroscopic they consist
of many different objects being the  elements of a structure
formation.

In connection with the absence of any standard view on the interaction
mechanisms  of  weak and super-weak external fields with  biological
systems and, especially with "a problem $k_BT$", in the papers \cite{Zakirjon} %[8]
 we proposed a simplified model to study an influence of the weak
 external magnetic and electric fields on the fluctuation of a stochastic
 ionic current in blood vessels and leading to a new method of solution
 of the $k_BT$ problem in magnetobiology. The present work is written
 on the basis of results of works \cite{Zakirjon}% [8]
extending the analysis performed there.

The aim of the present work is to estimate energy of molecules near
to resonant value of a magnetic field.

\section{The Langevin Equation and its Solution}
\subsection{Basic Equation}

More than 90\% of biological tissues  consists of polar molecules of fibers,
nucleinic acids, lipids, fats, carbohydrates and water. The blood of the
human being is the multi-component system consisting of plasma and blood
cells. As it is known from the physiology of cardiovascular system of
people \cite{Marmon}, %[9],
plasma of blood is a water solution of electrolytes, nutrients,
metabolites, fibers and vitamins. The electrolytes structure of plasma
reminds the sea water
that is connected with the evolution of  a life form from the sea.
Concentrations of ions like Na$^+$, Ca$^{2+}$ and Cl$^-$   in plasma
of blood are larger than in cytoplasm.  On the contrary, the concentration
of ions K$^+$, Mg$^{2+}$ and phosphate in plasma of blood are lower
than one in cells. These facts allow us to consider the blood in blood
vessels as a conductor of an ionic electric current.

It is well known that in all conductors the fluctuations of a current
take place because of their molecular structure. Such effect has been
experimentally measured by Johnson in 1928 and it denoted as the
Johnson noise \cite{Johnson}. %[10].
The spectral density of the Johnson noise does not depend on a frequency
and, therefore, it represents the white electric noise. The Johnson noise
is observed in the systems  being in the equilibrium states or close to them.

The concrete microscopic mechanisms for the occurrence of the Johnson noise
can be different. However in all cases the Johnson noise is caused by the
chaotic Brownian motion of the charged particles which possesses two
important properties: fast casual change of the direction and the basic
opportunity of carrying a large charge through the section of a conductor.
Thus the geometrical shape of the system considered is of no relevance for
the process. The Brownian  character of the charge carriers' motion remains
the same. As the Brownian  motion of ions is very poorly connected with
fluctuations of their number:  a disappearance of one particles and a
creation of others does not change the  essence of the process analyzed.
An arbitrarily large number of charge can be transferred by any way through
the given section inside of a conductor \cite{Bochkov,Keizer}. %[11,12].

This process is stationary, Gaussian and Markovian process. The
random ionic current satisfies a linear stochastic differential
equation, namely the Langevin equation. An element of the vascular
system with a weak random current is assumed to be described as a
line element of random current with the length $L$  located in a
weak external static magnetic field ($B$). As it is well-known, in
this case the acting force  onto the element of random current is
the force $F(t)=i(t)LB\sin(\alpha)$, where $\alpha$ is between
directions of the current element  $\overrightarrow{L}$  and
magnetic field $\overrightarrow{B}$ \cite{Edward}.% [13].

In the subsequent analysis we consider the fluctuations of a
scalar quantity, the magnitude of the random current. The mentioned
circumstances allow us to formulate the basic equations in the
following form \cite{Zakirjon} %[8]
\begin{equation}
\label{dit} % Eq.1
\frac{di(t)}{dt}=-\Lambda i(t)+f(t).
\end{equation}
Here $\Lambda=\lambda-\frac{qn_{ch}}{m}B\sin\alpha$, where $m$,
$q$, $n_{ch}$ are the  mass,  charge and number of ions, respectively,
in the volume $V$; $B$ is the induction of an external magnetic field,
\begin{equation}
\label{fric} % Eq.2
\lambda=\frac{k_BT}{mD}
\end{equation}
 is the friction coefficient,
\begin{equation}
\label{diff} % Eq.3
D=\frac{k_BT}{6\pi\eta r}
\end{equation}
is the diffusion coefficient,  $r$-ionic radius, $\eta$ is the
viscosity coefficient of liquid;
\begin{equation}
\label{diff} % Eq.4
f(t)=\frac{q}{mL}\sum_i f_i(t),
\end{equation}
where $f_i(t)$ is the random force acting on the corresponding
particle. It is the same for any atom of the same type and it is not
correlated with the random forces acting on other type ions
\cite{Keizer,Kampen}:
\begin{equation}
\label{ft0}  % Eq.5
<f(t)>=0, \,\,<f(t)f(t')>=\gamma \delta(t-t'),
\end{equation}
where
\begin{equation}
\label{gamma}  % Eq.6
\gamma=\frac{2k_BTq^2n_{ch}\lambda}{mL^2}
\end{equation}
is the intensity of the Langevin source.

\subsection{Solution of the Basic Equation and General Analysis}

As shown above, the random electric current may be described by the
linear stochastic differential equation with white noise as the
random source. The process under consideration is a process of
Ornstein-Uhlenbeck type and the formal solution of Eq.(\ref{dit})
may be presented in the form \cite{Zakirjon,Keizer}
\begin{equation}
\label{it} % Eq.7
i(t)=e^{-\Lambda t}i(0)+\int_0^t e^{-\Lambda (t-\tau)}f(\tau)d\tau .
\end{equation}
The average value of the current may be determined from Eqs. (\ref{ft0})
and (\ref{it}) as
\begin{equation}
\label{averit} % Eq.8
<i(t)>=e^{-\Lambda t}<i(0)>.
\end{equation}

One can also easily compute the dispersion of the random current fluctuations, i.e.
$\sigma(t)=<i^2(t)>-<i(t)>^2$. This quantity is given by
\begin{equation}
\label{sigmat}  % Eq.9
\sigma(t)=e^{-2\Lambda t}\int_0^t e^{2\Lambda \tau}\gamma d\tau .
\end{equation}

Taking the derivative of Eq.(\ref{sigmat}) with respect to time, we obtain
\begin{equation}
\label{dsigmat}  % Eq.10
\frac{d\sigma(t)}{dt}=-2\Lambda \sigma(t)+\gamma.
\end{equation}

The solution of Eq. (\ref{dsigmat}) with the initial condition
$\sigma(0)=0$ has the form
\begin{equation}
\label{sigmatin} % Eq.11
\sigma(t)=\sigma(\infty) (1-e^{-2\Lambda t}),
\end{equation}
where
\begin{equation}
\label{sigmainf}  % Eq.12
\sigma(\infty)=\lim_{t\rightarrow\infty}\sigma(t)=\frac{\gamma}{2\left(\lambda-
\frac{qn_{ch}B\sin\alpha}{m}\right)}.
\end{equation}
    As evident from Eq. (\ref{sigmainf}), the fluctuations of the random
    ionic electric current have a resonant character at
\begin{equation}
\label{lambda}  % Eq. 13
\lambda=\frac{qn_{ch}B\sin\alpha}{m}.
\end{equation}
The corresponding magnetic induction $B$ is determined by expression
\begin{equation}
\label{Binduc}  % Eq.14
B=\frac{\lambda m}{qn_{ch}\sin\alpha}.
\end{equation}

\section{Resonant energy of molecules}
\subsection{Basic formulas}

For any stochastic process $i(t)$  the power spectrum $i^2(t))$  is
defined as a function
of  the spectral density $S(\omega)$ by the relation \cite{Keizer,Kampen}% [12,14]
\begin{equation}
\label{iit}  % Eq.15
<i^2(t)>=\frac{1}{2\pi}\int_{-\infty}^{\infty}S(\omega)d\omega.
\end{equation}
For a scalar process with the real part of the relaxation rate  $\Lambda$,
the spectral density has the following form
\begin{equation}
\label{S}    % Eq.16
S(\omega)=\frac{\gamma}{\omega^2+\Lambda^2}.
\end{equation}
For the fluctuation of an random ionic current the spectral density is
connected with the power $P$ disseminated by the current at the given frequency as
\begin{equation}   %Eq.17
\label{Pin}
P=\frac{1}{\pi}\int_{0}^{\infty} R S(\omega)d\omega,
\end{equation}
where
\begin{equation}
\label{R}% Eq.18
R=\frac{m L^2\lambda}{n_{ch}q^2}
\end{equation}
is the electric resistance of a considered element of current  with a length $L$
\cite{Keizer}. %[12].
We substitute (\ref{S}) in (\ref{Pin}) after integrating over $\omega$, we obtain
\begin{equation}
\label{P} % Eq.19
P=\frac{R\gamma}{2\Lambda},
\end{equation}
 We substitute (\ref{R}) into (\ref{P}) and the resulting expression we
 multiply with  the exposition time $t$ then we divide it by the total number
 of molecules in the  considered volume $V$, i.e.
 $n_{tot}\approx N \cdot V$ ($N\approx 10^{28}m^{-3}$).
 In this way, we obtain the average energy of a molecule \cite{Zakirjon} %[8]
\begin{equation}
\label{enermol} % Eq.20
\varepsilon=\frac{Pt}{n_{tot}}=\frac{k_B T \lambda^2}{n_{tot}\Lambda}t .
\end{equation}

Because $i(t)$  is a stationary Gaussian process, Eqs.(\ref{it}) and
(\ref{sigmatin}) are sufficient  to completely determine the  conditional
density of probability $P_2$. It is taken from  \cite{Keizer,Kampen} %[12,14]
\begin{equation}    %Eq. 21
\label{p2}
P_2(i(0)\mid i(t),t)=\frac{1}{\sqrt{2\pi\sigma(t)}}\exp\left[-\frac{(i(t)-i(0)
\exp(-\Lambda t))^2}{2\sigma(t)}\right].
\end{equation}

One can see that the width of the conditional probability distribution depends
on $\sigma(t)$  and at the large values of $\sigma(t)$  the density of
probability goes to zero.

\subsection{Numerical Estimations}

In order to estimate the resonant value of the magnetic induction as described
by Eq.(\ref{Binduc}), we employ the following data \cite{Marmon}: %[9]:
for a person of 70 kg weight, we have an amount of 1.7 kg calcium, 0.25 kg
potassium, 0.07 kg sodium, 0.042 magnesium, 0.005 kg iron, 0.003 kg zinc.
The effect of calcium in the organism of a human being is very significant.
Its salts are a permanent constituent of the blood, of the cell and tissue fluids.
Calcium is a component part of the cell nucleus and plays a major role in the
processes of cell growth. 99\% of the calcium is concentrated in the bones,
the remaining part in the blood system and tissues.

The blood composes about 8.6\% of the mass of a human body. Hereby the fraction
of the blood located in the arteries is lower than 10\% of its total amount.
The same amount of blood is contained in the veins, the remaining 80\% are
contained in smaller units like the microvasculature, arterioles, venues and
capillaries. The typical values of the viscosity of blood plasma of a healthy
human being at $37^{\circ}$C are $1.2\cdot 10^{-3}$ Pa$\cdot$s \cite{Marmon}. %[9]
 The density of the blood is of the order $\rho$= (1.06--1.064) $\cdot 10^3$
 kg/m$^3$ \cite{Marmon}. %[9].
 Knowing the radius of the ions, we may determine the diffusion coefficient which
 is estimated as $D$= (1.8--2.0) $\cdot 10^{-9}$ m$^2$/s. The
 friction coefficient is calculated by formula $\lambda=6 \pi \eta r/m$ obtained
 from formulas (\ref{fric}) and (\ref{diff}). Its value has been obtained:
  $\lambda$=(3--6)$\cdot 10^{13}$  s$^{-1}$. For example, for  calcium ions
  we used $r_{\rm Ca}=10^{-10}$m \cite{Shannon},  $m_{\rm Ca}=6.68 10^{-26}$
  kg and $q_{\rm Ca}=2 \cdot 1.6  \cdot 10^{-19}$ C and we have obtained
  $\lambda =3.52 \cdot 10^{13}$
 s$^{-1}$. The aorta can be considered as a canal with a diameter of
 (1.6--3.2) $\cdot 10^{-2}$ m and a cross section area of
 (2.0--3.5) $\cdot 10^{-4}$ m$^2$, which splits of step by step into a
 network of $10^9$  capillaries each of them having a cross section area of
 about $7.01\cdot 10^{-12}$ m$^2$ with an average length of about 10$^{-3}$ m.

The number of calcium ions in a volume $V$=(2--3.5)$\cdot 10^{-6}$ m$^3$ of
the aorta is equal to $n_{ch}$=(0.8--1.4)$\cdot 10^{19}$, in a volume
$V=7\cdot 10^{-15}$ m$^3$ of the capillary we have  $n_{ch}=2.7\cdot 10^{10}$.
Substituting these values into Eq.(\ref{Binduc}), we get, at $\sin(\alpha)\approx 1$,
for the aorta $B\approx 0.5\cdot 10^{-12}$ T and for the capillary $B\approx 270$ $\mu$T.

For numerical estimates average energy of a molecule, we express the parameter
in the following form
\begin{equation}
\label{LambdaL} % Eq.22
\Lambda=\lambda \left(1-\frac{\omega(B)}{\lambda}\right),\,\,{\rm where}\,\,
\omega(B)=\frac{qn_{ch}}{m}B.
\end{equation}
Substituting into Eq.(\ref{enermol}) the values  of the total number of molecule
 in a volume $V=7\cdot 10^{-15}$ m$^3$  for capillary $n_{tot}\approx 10^{13}$
 and  total number of molecule in a volume $V$=(2--3.5)$\cdot 10^{-6}$m$^3$
 for the aorta $n_{tot}\approx 10^{22}$.

 The  energy received  by a molecule in a  capillary during $t=1$s was calculated
 for the $\Lambda=0.5\lambda$ , 0.05$\lambda$, and  0.005$\lambda$ which correspond
 to values of the induction of an external magnetic field
$B$=135 $\mu$T, 256.5$\mu$T  and 268.65$\mu$T, respectively, because
as it was mentioned above $B$=270$\mu$T is the resonant value for a
capillary. At these values of $B$ we obtain the following  estimates for the
energy received by a molecule in a  capillary during $t=1$s:
 $\varepsilon\approx 2k_B T$, $\varepsilon\approx 20k_B T$ and
 $\varepsilon\approx 200k_B T$.
 The similar estimations for the $\Lambda=0.5\lambda$ , 0.05$\lambda$,
 and  0.005$\lambda$ for a molecule in the aorta led us to values
 $\varepsilon\approx 2\cdot 10^{-9}k_B T$, $\varepsilon\approx 20\cdot 10^{-9}k_B T$
 and $\varepsilon\approx 200\cdot 10^{-9}k_B T$. Apparently,
 it follows from these estimates that under identical conditions a molecule
 of the aorta has $10^{-9}$  times less energy than the molecules of capillaries.

 At $\Lambda=0.5\cdot 10^{-9}\lambda$  for the molecule of the aorta we get:
 $\varepsilon\approx 2k_B T$, but according to Eq.(\ref{p2}) the probability of
 such a process is close to zero.  These estimates show that large vasculatures
 are more sensitive to ultra-weak field and capillaries are sensitive to weak
 and moderate magnetic fields.

\section{Conclusions}

The numerical estimations showed that the resonant values of the energy of
molecular motion in the capillaries and aorta are different. These estimations
proved further that under identical conditions a molecule of the aorta gets
10$^{-9}$ times less energy than the molecules of the capillaries.
The capillaries are very sensitive to the resonant effect, with an approach
to the resonant value of the magnetic field strength, the average energy of
the molecule localized in the capillary  increases by several orders of magnitude
as compared to its thermal energy, this value of the energy is sufficient for
the deterioration of the chemical bonds. Even if the magnetic field has values
not so near to the resonant values, with an increase of the time of exposition
to the magnetic field a significant effect can be reached.
A series of experiments are desirable to check the suggested mechanism  of
an action of the weak magnetic fields on the biological objects, especially,
``a $k_BT$ problem''.

\begin{acknowledgments}
Authors thank Drs. G. G. Adamian and N. V. Anfonenko for valuable
discussions and comments. Z. Kanokov is grateful to
 the Deutsche Forschungsgemeinschaft
(DFG 436 RUS 113/705/0-3) for the financial support.

\end{acknowledgments}

\end{document}